\documentclass[cameraready]{Interspeech}
\title{VoCodec: A Low-bitrate Streamable Neural Speech Codec with Voicing-driven Quantization}

\author[affiliation={1}]{Xiao-Hang}{Jiang}
\author[affiliation={1}, correspondingauthor]{Yang}{Ai}
\author[affiliation={1}]{Rui-Chen}{Zheng}
\author[affiliation={1}]{Li-Rong}{Dai}
\author[affiliation={1}]{Zhen-Hua}{Ling}
\author[affiliation={2}]{Ji}{Wu}
\address{
    $^1$ University of Science and Technology of China, China \\
    $^2$ Tsinghua University, China
}
\email{
\{jiang\_xiaohang, zhengruichen\}@mail.ustc.edu.cn, \\
\{yangai, lrdai, zhling\}@ustc.edu.cn, wuji\_ee@tsinghua.edu.cn
}

\usepackage{comment}
\usepackage{multirow}
\begin{document}

\maketitle

\addtolength{\textfloatsep}{-0.2cm}
\addtolength{\dbltextfloatsep}{-0.2cm}

\begin{abstract}
% Neural speech codecs, as key components for compressing and reconstructing speech signals, play a significant role in speech transmission and storage. 
% However, most existing codecs employ a uniform quantization strategy across all speech frames, allocating the same bitrate regardless of content. 
% This approach is suboptimal for speech signals, resulting in unnecessary bitrate consumption and leaving potential for further compression. 
% In this paper, we present a low-bitrate streamable neural speech codec called VoCodec. 
% Unlike existing codecs, VoCodec employs a voicing-driven quantization strategy, assigning different bitrates to voiced and unvoiced frames based on their sensitivity to human auditory perception. 
% Specifically, VoCodec incorporates a voicing detector into its fully causal encoder–quantizer–decoder neural coding framework to identify voicing characteristics in the input speech. 
% Based on this, it adopts complex residual scalar-vector quantization for voiced frames and simple scalar quantization for unvoiced frames during quantization. 
% Experiments show that on the LibriTTS dataset at a 16 kHz sampling rate, VoCodec outperforms baseline neural speech codecs even at a bitrate as low as 1.1 kbps. 
% Our further experiments also confirm that introducing voicing-driven quantization can effectively reduce the bitrate by approximately 27\% compared to the original uniform quantization strategy.

Neural speech codecs are key to speech transmission and storage, but most use uniform quantization across frames, allocating the same bitrate regardless of content and wasting bits. We propose VoCodec, a low-bitrate streamable neural speech codec with voicing-driven quantization that assigns higher bitrate to voiced frames and lower bitrate to unvoiced frames according to perceptual sensitivity. VoCodec embeds a voicing detector in a fully causal encoder-quantizer-decoder neural coding framework, using residual scalar-vector quantization for voiced frames and simple scalar quantization for unvoiced ones. Experiments show that on the LibriTTS dataset at a 16 kHz sampling rate, VoCodec outperforms baseline neural speech codecs even at a bitrate as low as 1.1 kbps. Our further experiments also confirm that introducing voicing-driven quantization can effectively reduce the bitrate by approximately 27\% compared with uniform quantization strategy.

\end{abstract}

\section{Introduction}
\label{sec:intro}
Speech codec is a critical component in digital speech processing, serving the dual functions of encoding and decoding speech signals. They play a vital role in reducing the data volume required to represent speech while maintaining acceptable decoded speech quality. These codecs find wide application in speech communication \cite{salami1994toll}, speech compression \cite{brandenburg1994iso}, and numerous downstream tasks such as speech synthesis \cite{wang2023neural,borsos2023audiolm,mehta2024matcha,kim2021conditional} and speech enhancement \cite{pascual2017segan,xue2024low}.

Traditional speech codecs, such as code-excited linear prediction (CELP) \cite{schroeder1985code}, leverage the short-term stationarity of speech to model vocal tract resonances via linear predictive coefficients (LPC) \cite{o1988linear}. By distinguishing between voiced and unvoiced excitation sources, CELP achieves high-quality reconstruction at medium bitrates (e.g., 4–16 kbps). However, these methods are difficult to generalize to different speech features and often rely on manual parameter adjustments.

% As deep learning continues to advance, neural speech codecs are increasingly becoming a central focus in speech coding research.
% In the early stage, SoundStream \cite{zeghidour2021soundstream} and Encodec \cite{defossez2023high} were proposed. 
% Compared with traditional codecs, they achieved significant performance improvements while maintaining the characteristic of low latency, that is, using causal models that do not require future information input. 
% However, this also led to significant room for improvement in their coding quality. 
With advances in deep learning, neural speech codecs have emerged as a key focus in speech coding research. Early works like SoundStream \cite{zeghidour2021soundstream} and Encodec \cite{defossez2023high} achieved marked performance gains over traditional codecs while maintaining low latency via causal models (no need for future input). However, this left substantial room for improving coding quality.
Later, AudioDec \cite{wu2023audiodec} was proposed, which attempted to use HiFi-GAN \cite{kong2020hifi} vocoder as a decoder to improve coding quality while maintaining low latency, but its performance was still unsatisfactory. 
As the demand for higher quality increased, non-causal codecs that ignore latency requirements gradually emerged \cite{kumar2024high,liu2024semanticodec,xin2024bigcodec}. 
By scaling up model parameters, they achieved impressive performance; however, this limited their applicability in real-world scenarios such as real-time speech communication.
To meet the dual objectives of low latency and high perceptual quality, StreamCodec \cite{jiang2025streamable} adopts a fully causal and lightweight architecture, takes the modified discrete cosine transform (MDCT) spectrum as the modeling target, and employs a novel residual scalar-vector quantizer (RSVQ) for discretization. 
The RSVQ adopts a coarse-to-fine quantization strategy, which significantly improves coding quality.

However, most neural speech codecs use uniform quantization, allocating the same bitrate to all frames regardless of their characteristics (e.g., voiced or unvoiced). Since voiced frames, with their periodic structure and concentrated energy, contribute more to speech intelligibility, while unvoiced frames have a weaker perceptual impact \cite{knorr1979reliable}, uniform allocation wastes bits on unvoiced frames and leaves room for bitrate reduction.

To address this issue, we propose VoCodec, a low-bitrate streamable neural speech codec with a voicing-driven quantization strategy. Inspired by StreamCodec \cite{jiang2025streamable}, VoCodec adopts a fully causal architecture for streaming generation. It first detects the voicing attribute of each frame, then applies RSVQ to voiced frames and a simple scalar quantizer (SQ) to unvoiced frames \cite{mentzer2024finite}. Experimental results show that, compared with baseline models using uniform quantization, VoCodec reduces bitrate by approximately 27\% while maintaining high reconstruction quality by allocating more bits to perceptually important voiced components.
\begin{figure*}
    \centering
    \includegraphics[width=0.92\linewidth]{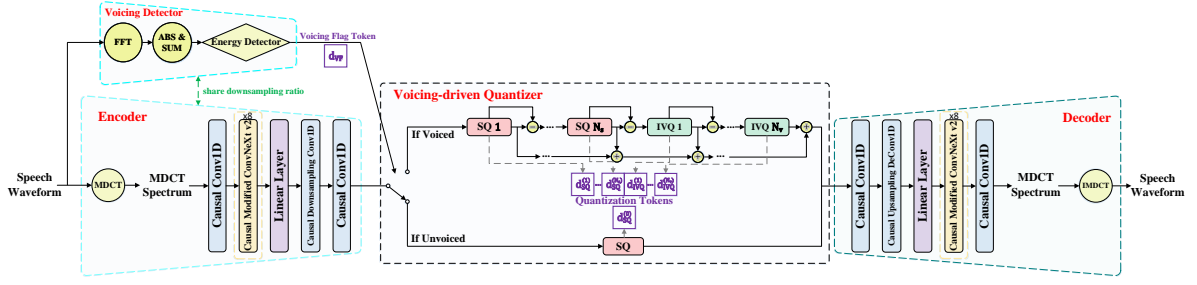}
    \caption{Overall architecture of the proposed VoCodec. Here, MDCT, IMDCT, Uni-LSTM, SQ, IVQ, FFT, ABS and SUM stand for modified discrete cosine transform, inverse modified discrete cosine transform, unidirectional long short-term memory layer, scalar quantizer, improved vector quantizer, fast Fourier transform, absolute value calculation and summation, respectively.
    }
    \label{overview}
\end{figure*}

\section{Proposed Method}
\label{sec: propose}
\subsection{Overview}
Fig. \ref{overview} shows an overview of the proposed VoCodec.
VoCodec consists of four main components: an encoder, a voicing detector, a voicing-driven quantizer and a decoder. 
The encoder and the voicing detector process the input speech in parallel, and their outputs share the same frame rate (i.e., they share the same downsampling ratio). 
The voicing-driven quantizer receives both the encoder output and the voicing flag token from the voicing detector, and applies either RSVQ or SQ to the encoded features based on the detected voicing flags.
Finally, the decoder decodes the quantized features and reconstructs the speech waveform.

\subsection{Encoder \& Decoder}

As shown in Fig. \ref{overview}, VoCodec inherits StreamCodec’s \cite{jiang2025streamable} fully causal convolutional encoder-decoder. 
Its encoder encodes the MDCT spectrum extracted from input speech, using 8 convolution-based causal modified ConvNeXt v2 blocks \cite{jiang2025streamable} as backbone, along with causal convolutional layers and linear layers.
As a supplement to the causal convolution operations, we introduce a unidirectional long short-term memory (LSTM) layer \cite{hochreiter1997long} at the end of the encoder for sequential modeling.
% Supplemented by causal conv layers (for feature dim adjustment) and a downsampling layer, its final layer uses a unidirectional LSTM \cite{hochreiter1997long} for sequential modeling. 
The decoder, symmetric to the encoder but with upsampling instead of downsampling, decodes the MDCT spectrum which is finally converted to waveform via inverse MDCT.
% As shown in Fig. \ref{overview}, VoCodec refers to the model structure of StreamCodec \cite{jiang2025streamable} that both the encoder and decoder of VoCodec employ fully causal convolutional networks. 
% For the encoder, the MDCT spectrum is first extracted from the input speech and then fed into the network for processing. 
% The modified ConvNeXt v2 network \cite{ai2024apcodec} serves as the backbone of the encoder network, and its residual connection structure includes a 1D depthwise convolutional layer, a layer normalization, a linear layer and a Gaussian error linear unit (GELU) activation. 
% In addition to the modified ConvNeXt v2 network, two 1D causal convolutional layers are deployed at the start and end of the encoder to adjust the dimensionality of the features respectively, and a 1D causal downsampling layer is used to compress the features. At the final layer of the encoder, we adopt a unidirectional LSTM network to enhance the model's sequential modeling capabilities.
% The decoder and encoder are symmetric in structure, with the only difference being that the decoder replaces the downsampling layer in the encoder with the upsampling layer. 
% The decoder outputs the decoded MDCT spectrum, which is subsequently transformed into the speech waveform via inverse MDCT (IMDCT).

\subsection{Voicing Detector}

In VoCodec, the voicing detector extracts voicing flag tokens from the input speech waveform. Based on these tokens, the voicing-driven quantizer adaptively selects quantization strategies to optimize bitrate allocation. Given that VoCodec operates in a streamable manner, we illustrate the processing pipeline using a single frame as an example. The process for generating the voicing flag token is detailed as follows.
% In VoCodec, the voicing detector extracts voicing flag tokens from the input speech waveform. 
% The voicing-driven quantizer selects different quantization strategies based on the flag tokens to achieve more efficient bitrate allocation.
% Given that VoCodec operates in a streamable manner, we illustrate the processing pipeline using a single frame as an example.
% The process for generating the voicing flag token is detailed as follows. 
First, for a frame of time-domain speech waveform $\bm{x}\in\mathbb R^N$ (i.e., a speech segment after framing and windowing), it first computes the complex spectrum using the fast Fourier transform (FFT), i.e.,
\begin{equation}
\begin{aligned}
\bm{X}[k] = \sum_{n=0}^{N-1} \bm{x}[n] \cdot e^{-j2\pi kn/N},
\end{aligned}
\end{equation}
where \(N\) and \(k\) represent frame length and frequency bin index, respectively. 
To align the frame rate of the voicing flag token with that of the encoder’s encoded feature, the framing process adopts a frame shift identical to the encoder’s downsampling ratio.
Subsequently, the frame energy is calculated according to the formula:
\begin{equation}
\begin{aligned}
E = \sum_{k=K_{f0min}-1}^{K_{f0max}-1} |\bm{X}[k]|,
\end{aligned}
\end{equation}
where \(K_{f0\text{min}}\) and \(K_{f0\text{max}}\) are the index boundaries of the fundamental frequency search range in the frequency domain, defined as
$K_{f0\text{min}} = \left\lfloor \frac{f_{0{\text{min}}} \cdot K_{\text{FFT}}}{f_s} \right\rfloor, 
K_{f0\text{max}} = \left\lfloor \frac{f_{0{\text{max}}} \cdot K_{\text{FFT}}}{f_s} \right\rfloor$,
where \(f_{0{\text{min}}}\) and \(f_{0{\text{max}}}\) are minimum and maximum fundamental frequency. 
\(K_{\text{FFT}}\) and \(f_s\) represent the number of FFT points and the waveform sampling rate, respectively.
Next, $E$ is processed by an energy detector. 
If the energy exceeds the threshold $\tau_{\text{energy}}$, the frame is classified as voiced (assigned a flag token of 1); otherwise, it is classified as unvoiced (assigned a flag token of 0), i.e.,
\begin{equation}
\begin{aligned}
d_{VF} = \begin{cases} 1 & \text{if } E > \tau_{energy} \\ 0 & \text{otherwise} \end{cases},
\end{aligned}
\end{equation}
where $d_{VF}$ is the voicing flag token.

\subsection{Voicing-driven Quantizer}

The voicing-driven quantizer is a core component of VoCodec, specifically designed to apply distinct quantization strategies to voiced and unvoiced frames.
Due to the periodic nature and low-frequency energy concentration of voiced sounds, they play a more dominant role in auditory perception.
In contrast, unvoiced sounds exhibit more dispersed and lower energy, resulting in a relatively minor perceptual impact. 
Therefore, a more sophisticated quantization strategy is applied to voiced frames, as they contribute more significantly to perceived speech quality than unvoiced frames.

% The voicing-driven quantizer, a core component of VoCodec, applies distinct quantization strategies to voiced and unvoiced frames. Voiced sounds, featuring periodicity and concentrated low-frequency energy, dominate auditory perception, whereas unvoiced sounds have dispersed, lower energy and a minor perceptual impact. Thus, a more sophisticated quantization is applied to voiced frames, as they contribute more significantly to perceived speech quality than unvoiced frames.

Specifically, given an encoded feature frame $\bm{s} \in \mathbb{R}^M$ from the encoder, where $M$ is the feature dimension, the voicing-driven quantizer selects either RSVQ or SQ based on the voicing flag token $d_{VF}$. 
If \(d_{VF} = 1\) (i.e., $\bm{s}$ is the encoded feature of a voiced frame), $\bm{s}$ is quantized using the RSVQ. 
The RSVQ consists of multiple SQs \cite{mentzer2024finite} and improved vector quantizers (IVQs) \cite{zheng2025ervq}. 
The SQ handles coarse quantization, while IVQ is responsible for fine quantization, with a residual connection established between them. 
In contrast to traditional VQ, IVQ introduces online clustering and codebook balancing loss. These improvements enable IVQ to prevent codebook collapse and achieve higher codebook efficiency. 
If \(d_{VF} = 0\) (i.e., $\bm{s}$ is the encoded feature of an unvoiced frame), $\bm{s}$ is quantized using a single SQ. 
Therefore, the quantization result of $\bm{s}$ by the voicing-driven quantizer is:

\begin{equation}
\scalebox{0.88}{$
\hat{\bm{s}}=
\begin{cases}
\sum_{i = 1}^{N_s} \mathbb{L}(\mathbb{W}_{SQ}^{(i)},d_{SQ}^{(i)}) + \sum_{j = 1}^{N_v} \mathbb{L}(\mathbb{W}_{IVQ}^{(j)},d_{IVQ}^{(j)}), & \text{if } d_{VF} = 1, \\[4pt]
\mathbb{L}(\mathbb{W}_{SQ}^{(0)},d_{SQ}^{(0)}), & \text{if } d_{VF} = 0.
\end{cases}
$}
\end{equation}

% \begin{equation}
% \hat{\bm{s}}=
% \begin{cases}
% \begin{aligned}
% &\sum_{i=1}^{N_s} \mathbb{L}\!(\mathbb{W}_{SQ}^{(i)}, d_{SQ}^{(i)})
% \\
% &\quad + \sum_{j=1}^{N_v} \mathbb{L}\!(\mathbb{W}_{IVQ}^{(j)}, d_{IVQ}^{(j)}),
% \end{aligned}
% & d_{VF}=1, \\[2pt]
% \mathbb{L}\!(\mathbb{W}_{SQ}^{(0)}, d_{SQ}^{(0)}),
% & d_{VF}=0.
% \end{cases}
% \end{equation}

where $\hat{\bm{s}} \in \mathbb{R}^M$.
\(N_s\), \(N_v\) are the numbers of SQs and IVQs in RSVQ. 
$\mathbb{L}$ represents the lookup operation. 
$\mathbb{W}_{SQ}^{(i)}$ and $d_{SQ}^{(i)}$ denote the codebook (with a size of $K_{SQ}^{(i)}$) and the quantization token of the $i$-th SQ in RSVQ, respectively. 
$\mathbb{W}_{IVQ}^{(j)}$ and $d_{IVQ}^{(j)}$ denote the codebook (with a size of $K_{IVQ}^{(j)}$) and the quantization token of the $j$-th IVQ in RSVQ, respectively. 
$\mathbb{W}_{SQ}^{(0)}$ and $d_{SQ}^{(0)}$ denote the codebook (with a size of $K_{SQ}^{(0)}$) and the quantization token of the single SQ, respectively. 
Therefore, the quantization tokens produced by the voicing-driven quantizer are given as follows:
% \(\mathbb{L}\) denotes the lookup operation. For RSVQ, \(\mathbb{W}_{SQ}^{(i)}\) and \(d_{SQ}^{(i)}\) represent the codebook (size \(K_{SQ}^{(i)}\)) and quantization token of the i-th SQ, respectively; \(\mathbb{W}_{IVQ}^{(j)}\) and \(d_{IVQ}^{(j)}\) denote the codebook (size \(K_{IVQ}^{(j)}\)) and quantization token of the j-th IVQ, respectively. For the single SQ, \(\mathbb{W}_{SQ}^{(0)}\) and \(d_{SQ}^{(0)}\) are its codebook (size \(K_{SQ}^{(0)}\)) and quantization token. The quantization tokens generated by the voicing-driven quantizer are given as follows:
\begin{equation}
\begin{aligned}
\bm{d}_Q=
\begin{cases}
d_{SQ}^{(1)},\dots,d_{SQ}^{(N_s)},d_{IVQ}^{(1)},\dots,d_{IVQ}^{(N_v)}, &\text{if }d_{VF}=1,\\
d_{SQ}^{(0)}, &\text{if }d_{VF}=0.
\end{cases}
\end{aligned}
\end{equation}
% RSVQ generates token $ \bm{T}_v[m] =d_{SQ}^{(1)},\dots,d_{SQ}^{(N_s)},d_{IVQ}^{(1)},\dots,d_{IVQ}^{(N_v)}$. SQ generates token $ \bm{T}_u[m] = d_{SQ}^{(0)}$.

Therefore, VoCodec produces two types of tokens, i.e., quantization tokens $\bm{d}_Q$ and voicing flag token $d_{VF}$. 
Once converted into binary format, these two types of tokens are both transmitted or stored, allowing the receiver or decompression module to reconstruct the speech signal accordingly based on the voicing flag.
For bitrate calculation of VoCodec, assume that the downsampling rate of the encoder is $D$ and the ratio of voiced frames is $R$, the bitrate of VoCodec is calculated as follows: 
\begin{equation}
    \begin{aligned}
    \label{bitrate}
Bitrate=\frac{f_s}{D}\cdot [R\cdot( \sum\limits_{i=1}^{N_s}  \log_2K_{SQ}^{(i)} + \sum\limits_{j=1}^{N_v} \log_2K_{IVQ}^{(j)}) \\+ (1-R)\cdot( \log_2K_{SQ}^{(0)} )+1].
    \end{aligned}
\end{equation}
% where the value of $\bm{T}_{flag}$ is consistent with the number of frames $\frac{f_s}{w_s\cdot D}$ output by the encoder per second.

\subsection{Mask-based Efficient Training}

% When training VoCodec, the streaming mode can lead to a significant slowdown in training speed due to the inability to perform parallel computation.
% To accelerate the training process, we adopt a mask-based training strategy. 
% Specifically, instead of selecting between RSVQ and SQ for the encoded features $\bm{S}=[\bm{s}_1,\dots,\bm{s}_B]\in\mathbb R^{M\times B}$ of a training batch based on the voicing flag token, we perform dual-path quantization on $\bm{S}$ in parallel, where $\bm{s}_b\in\mathbb{R}^M(b=1,\dots,B)$ and $B$ is the number of frames in a training batch. 
% This produces two quantized outputs: $\hat{\bm{S}}_v\in\mathbb R^{M\times B}$ from RSVQ and $\hat{\bm{S}}_u\in\mathbb R^{M\times B}$ from the single SQ. 
% Assuming the voicing flag token vector for the training batch is $\bm{d}_{VF}=[d_{VF-1},\dots,d_{VF-B}]^\top$, we construct the corresponding mask matrix as 
During VoCodec's training, streaming mode slows training significantly as parallel computation is not feasible. 
To accelerate training, we use a mask-based strategy.
Specifically, instead of selecting RSVQ or SQ for batch encoded features \(\bm{S}=[\bm{s}_1,\dots,\bm{s}_B]\in\mathbb R^{M\times B}\) (with $\bm{s}_b\in\mathbb{R}^M(b=1,\dots,B)$ and $B$ as the number of frames) based on voicing flag tokens, we apply dual-path quantization to \(\bm{S}\) in parallel. 
This yields two quantized outputs: \(\hat{\bm{S}}_v\in\mathbb R^{M\times B}\) from the RSVQ and \(\hat{\bm{S}}_u\in\mathbb R^{M\times B}\) from the SQ. Given the batch voicing flag token vector \(\bm{d}_{VF}=[d_{VF-1},\dots,d_{VF-B}]^\top\in\mathbb R^B\), we construct the corresponding mask matrix as
\begin{equation}
    \begin{aligned} 
\textbf{mask}=[\underbrace{\bm{d}_{VF},\bm{d}_{VF},\dots,\bm{d}_{VF}}_{repeat~M~times}]^\top,
    \end{aligned}
\end{equation}
where $\textbf{mask}\in\mathbb R^{M\times B}$.
At the output of the voicing-driven quantizer, the decoder receives:
\begin{equation}
    \begin{aligned} 
\hat{\bm{S}}=\hat{\bm{S}}_v\odot\textbf{mask}+\hat{\bm{S}}_u\odot(\textbf{1}-\textbf{mask}),
    \end{aligned}
\end{equation}
%At this point, the $\bm{d}_{VF}$ is equivalent to a mask sequence. 
where $\odot$ represents the element-wise multiplication. 
The purpose of the mask matrix is to selectively mask either voiced or unvoiced frames.
This method facilitates efficient and parallelizable model training. 
Regarding the loss function, VoCodec borrows the formulation from \cite{jiang2025streamable}, which consists of spectral-level loss, codebook loss and generative adversarial loss.

\begin{table}[t]
\centering

    % \caption{Experimental results on decoded speech quality, efficiency and complexity evaluations for compared codecs at 1.1 kbps on the LibriTTS test set with a 16 kHz sampling rate. The \textbf{bold} and \underline{underline} numbers indicate optimal and sub-optimal results, respectively.}
    % \caption{Results on LibriTTS (16 kHz) at 1.1 kbps. The \textbf{bold} and \underline{underline} numbers indicate optimal and sub-optimal results, respectively.}
    \caption{Results on LibriTTS (16 kHz) at 1.1 kbps. The \textbf{bold} and \underline{underlined} numbers indicate optimal and sub-optimal results, respectively. The hidden-reference natural speech scored $93.21\pm1.74$, and the anchor scored $29.25\pm2.68$ in MUSHRA.}
    \huge
    \resizebox{\linewidth}{!}{
    \begin{tabular}{c | c| c c c c c c}
        \hline
        
        \hline

        \hline
        &\multirow{2}{*}{Streamable}& \multicolumn{5}{c}{Target Bitrate = 1.1 kbps}  \\
        \cline{3-8}
         && LSD (dB) $\downarrow$ & STOI $\uparrow$ & ViSQOL $\uparrow$& MUSHRA $\uparrow$ &FLOPs $\downarrow$ &Param.$\downarrow$\\ 
         \hline
         DAC &$\times$& 0.936 & 0.888 & 3.781 & 74.83$\pm$5.32 &55.53G& 73.87M\\
         BigCodec&$\times$& \textbf{0.888} & \underline{0.920} & \underline{4.086}& \textbf{77.40$\pm$5.07} &61.03G& 159.32M\\
         AudioDec &$\checkmark$& 0.988 & 0.698 & 3.617& 71.02$\pm$5.89 &26.32G& 24.41M \\
         
         MDCTCodec-S &$\checkmark$& 0.952 & 0.867 & 3.772 &65.37$\pm$7.71&\textbf{2.32G}& \textbf{6.75M} \\
         StreamCodec &$\checkmark$& 0.918 & 0.896 & 4.048 & 69.64$\pm$6.52 &\textbf{2.32G}& \textbf{6.75M} \\
         VoCodec &$\checkmark$& \underline{0.896} & \underline{0.916} & \textbf{4.115} & \underline{75.18$\pm$5.16} &\underline{2.62G}&\underline{9.31M}  \\
        % \hline
        % r-VoCodec &0.959&0.823&3.673\\
        \hline
        
        \hline

        \hline
    \end{tabular}}
\label{16k}
\end{table}
\section{Experiments}
\label{sec: Experiments}

\begin{table}[t]
\centering
    % \caption{Results on VCTK (48 kHz) at 2.7 kbps. The \textbf{bold} and \underline{underline} numbers indicate optimal and sub-optimal results, respectively.}
    \caption{Results on VCTK (48 kHz) at 2.7 kbps. The \textbf{bold} and \underline{underlined} numbers indicate optimal and sub-optimal results, respectively. The hidden-reference natural speech scored $95.08\pm1.33$, and the anchor scored $28.16\pm3.42$ in MUSHRA.}
    \resizebox{=0.90\linewidth}{!}{
    \begin{tabular}{c| c | c c c c }
        \hline
        
        \hline
        &\multirow{2}{*}{Streamable}& \multicolumn{3}{c}{Target Bitrate = 2.7 kbps}  \\
        \cline{3-6}
         && LSD (dB) $\downarrow$ & STOI $\uparrow$ & ViSQOL $\uparrow$ & MUSHRA $\uparrow$\\
         \hline
         DAC &$\times$& 0.870 & 0.842  & 3.580 & 81.22 $\pm$ 3.41 \\
         BigCodec &$\times$& \underline{0.850} & \textbf{0.880}  & 3.678 & \textbf{84.73} $\pm$ 2.27 \\
         AudioDec& $\checkmark$& 0.886 & 0.771  & 3.695 & 81.01  $\pm$ 3.37 \\

         MDCTCodec-S &$\checkmark$& 0.862& 0.836  & 3.714 & 79.30 $\pm$ 4.41 \\
         StreamCodec& $\checkmark$&0.855 & 0.842  & \underline{3.802}&  79.38 $\pm$  3.92 \\
         VoCodec &$\checkmark$& \textbf{0.847} & \underline{0.857} & \textbf{3.840}&  \underline{82.95 $\pm$ 2.38} \\

        \hline
        
        \hline
    \end{tabular}}
\label{48k}
\end{table}

\subsection{Experimental Setup}

% We conducted experiments\footnote{Speech samples are available at: \url{https://pb20000090.github.io/APSIPA2025/}.} on LibriTTS \cite{zen2019libritts} and VCTK \cite{veaux2017superseded} datasets. 

% The LibriTTS contains approximately 263 hours of speech, using train-clean-100 and train-clean-360 as training sets, dev clean and test clean as validation and test sets, respectively. 
% The speech in LibriTTS was downsampled from 24 kHz to 16 kHz for our experiments. 
% The VCTK contains approximately 43 hours of speech. 40936 utterances were selected as the training set, and 2937 utterances were selected as the test set.

We conducted experiments\footnote{Speech samples are available at: \url{https://pb20000090.github.io/VoCodec/}.} on the LibriTTS \cite{zen2019libritts} and VCTK \cite{veaux2017superseded} datasets. For LibriTTS (16 kHz sampling rate), we used train-clean-100 and train-clean-360 for training, dev-clean and test-clean for validation and testing. 
For VCTK (48 kHz sampling rate), 40,936 utterances served as the training set and 2,937 as the test set.

VoCodec inherits the model configuration and training parameters from \cite{jiang2025streamable}. The downsampling rate is set to $D=320$
For the RSVQ, we used one SQ and two IVQs. For all SQs and IVQs in VoCodec, the codebook size is set to 1024.
The input and output dimensions of each quantizer are set to 32. 
We set \(f_{0{\text{min}}}=60, f_{0{\text{max}}}=600\) and $\tau_{energy} = 0.75$ as voicing detection parameters. 

% The effective bitrate of VoCodec is content-dependent and varies with the voiced-frame ratio $R$. Voiced frames are encoded by RSVQ (one SQ + two IVQs), unvoiced frames by a single SQ, and a 1-bit voicing flag is transmitted per frame; thus the bitrate is given by (\ref{bitrate}). With $K=1024$, (\ref{bitrate}) simplifies to
% $
% \mathrm{Bitrate}=\frac{f_s}{320}\,(11+20R),
% $
% showing a linear dependence on $R$. This yields $\approx0.55$--$1.55$ kbps at $16$ kHz ($R\in[0,1]$) and scales proportionally with $f_s$ (e.g., $\approx1.65$--$4.65$ kbps at $48$ kHz). With $R=0.55$ on LibriTTS and $R=0.35$ on VCTK, the resulting average bitrates are $1.1$ kbps and $2.7$ kbps, respectively.
The effective bitrate of VoCodec is content-dependent and varies with the voiced-frame ratio $R$. Voiced frames are encoded by RSVQ (one SQ plus two IVQs), whereas unvoiced frames are encoded by a single SQ, with an additional 1-bit voicing flag transmitted for each frame. Equation (\ref{bitrate}) yields the simplified bitrate expression
$
\mathrm{Bitrate}=\frac{f_s}{320}(11+20R),
$
which shows that the bitrate depends linearly on $R$. Since $R\in[0,1]$, the bitrate ranges from $0.55$ to $1.55$ kbps at $16$~kHz; accordingly, at $48$~kHz, the bitrate ranges from $1.65$ to $4.65$ kbps. With $R=0.55$ on LibriTTS and $R=0.35$ on VCTK, the resulting average bitrates are $1.1$ kbps and $2.7$ kbps, respectively.

% \begin{table}[t]
% \centering
%     \caption{Results on VCTK (48 kHz) at 2.7 kbps. The \textbf{bold} and \underline{underline} numbers indicate optimal and sub-optimal results, respectively.}
%     \resizebox{=0.90\linewidth}{!}{
%     \begin{tabular}{c| c | c c c c }
%         \hline
        
%         \hline
%         &\multirow{2}{*}{Streamable}& \multicolumn{3}{c}{Target Bitrate = 2.7 kbps}  \\
%         \cline{3-6}
%          && LSD (dB) $\downarrow$ & STOI $\uparrow$ & ViSQOL $\uparrow$ & MUSHRA $\uparrow$\\
%          \hline
%          DAC &$\times$& 0.870 & 0.842  & 3.580 & 81.22 $\pm$ 1.41 \\
%          BigCodec &$\times$& \underline{0.850} & \textbf{0.880}  & 3.678 & \textbf{84.73 $\pm$ 0.91} \\
%          AudioDec& $\checkmark$& 0.886 & 0.771  & 3.695 & 81.01  $\pm$ 1.31 \\

%          MDCTCodec-S &$\checkmark$& 0.862& 0.836  & 3.714 & 79.30 $\pm$ 1.64 \\
%          StreamCodec& $\checkmark$&0.855 & 0.842  & \underline{3.802}&  79.38 $\pm$  1.53 \\
%          VoCodec &$\checkmark$& \textbf{0.847} & \underline{0.857} & \textbf{3.840}&  \underline{82.95 $\pm$ 1.05} \\

%         \hline
        
%         \hline
%     \end{tabular}}
% \label{48k}
% \end{table}
\begin{figure}[t]
  \centering
  \includegraphics[width=0.90\linewidth]{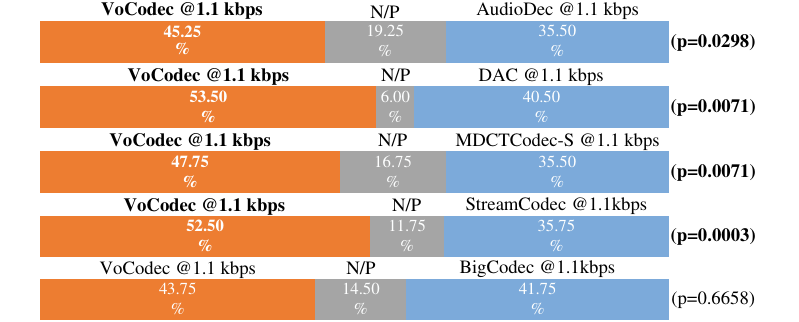}
  \caption{
  % Average preference scores (\%) of ABX tests comparing VoCodec and other codecs at 1.1 kbps on the LibriTTS test set (16 kHz). N/P denotes “no preference”, and $p$ is the paired $t$-test $p$-value.
  LibriTTS (16 kHz) ABX preference (\%) at 1.1 kbps. N/P indicates ``no preference”; $p$ is the paired $t$-test $p$-value.}
  \label{ABX_1}
\end{figure}

\begin{figure}[t]
  \centering
  \includegraphics[width=0.90\linewidth]{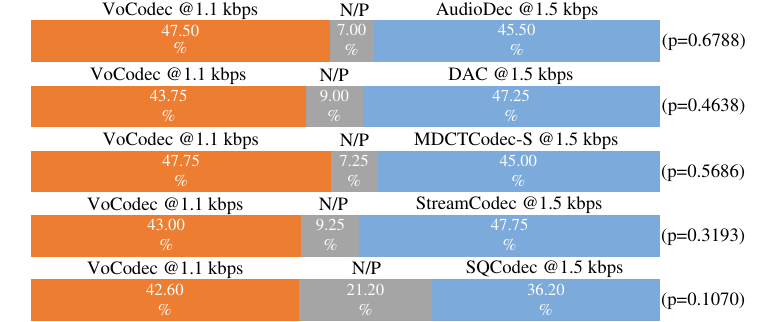}
  \caption{
  % Average preference scores (\%) of ABX tests comparing VoCodec at 1.1 kbps and other codecs at 1.5 kbps on the LibriTTS test set (16 kHz). N/P denotes ``no preference”, and $p$ is the paired $t$-test $p$-value.
LibriTTS (16 kHz) ABX preference (\%) comparing VoCodec (1.1 kbps) with other codecs (1.5 kbps).
N/P denotes “no preference”; $p$ is the paired $t$-test $p$-value.
  }
  \label{ABX_2}
\end{figure}

\subsection{Evaluation Metrics}

Both objective and subjective metrics are used for evaluating the performance of compared codecs from different perspectives. 
For coding quality evaluation, we employed three metrics: log-spectral distance (LSD), short-time objective intelligibility (STOI) \cite{taal2010short} and virtual speech quality objective listener (ViSQOL) \cite{chinen2020visqol}.
% , which respectively focused on the amplitude spectrum quality, intelligibility, and overall speech quality. 
Additionally, floating point operations (FLOPs) \cite{mcmahon1986livermore} and the number of model parameters (Param.) were also analyzed to assess computational complexity and model efficiency, respectively. 

% To assess subjective quality, we conducted ABX preference tests on the Amazon Mechanical Turk platform to compare our proposed VoCodec with other baseline codecs one by one.  
% In the ABX experiment, at least 20 native English-speaking listeners were asked to determine which utterance in each pair exhibited better speech quality or whether there was no preference. 
% Beyond computing the average preference scores, we also used the $p$-value from a $t$-test to assess the statistical significance of the difference between the two codecs being compared. 
For subjective evaluation, we conducted multiple stimuli with hidden reference and anchor (MUSHRA) \cite{recommendation2001method} and ABX preference tests on Amazon Mechanical Turk to compare VoCodec with baseline codecs.
% In MUSHRA, 20 test-set utterances per codec were rated by at least 20 native English listeners on a 0--100 scale, using natural speech as the reference (fixed at 100).
In MUSHRA, 20 test-set utterances per codec were rated by at least 20 native English listeners on a 0--100 scale, with natural speech as the hidden reference and a 3.5-kHz low-pass-filtered version as the anchor.
In ABX, 20 test-set utterance pairs for each comparison were evaluated by at least 20 native English listeners, who were asked to determine which utterance in each pair had better speech quality, or whether they had no preference.
% In ABX, 20 test-set utterances per codec were rated by at least 20 native English listeners, who chose A, B, or no preference.
We report mean scores/preferences and statistical significance via a $t$-test ($p$-value).

\subsection{Comparison with Baseline Neural Speech Codecs}

We compared the proposed VoCodec with several advanced neural speech codecs including non-streamable codecs, i.e., DAC \cite{kumar2024high}, BigCodec \cite{xin2024bigcodec}, SQCodec \cite{zhai2025one}, and streamable codecs, i.e., AudioDec \cite{wu2023audiodec}, causal MDCTCodec (denoted as MDCTCodec-S) \cite{jiang2024mdctcodec} and StreamCodec \cite{jiang2025streamable}. 
To ensure fairness, all baseline codecs were configured to match VoCodec’s target bitrates (1.1 kbps at 16 kHz and 2.7 kbps at 48 kHz). Specifically, we set the stride factors to 2, 4, 5, and 8, yielding an overall downsampling factor of 320, and used two codebooks with sizes of 2048 at 16 kHz and 512 at 48 kHz. SQCodec was excluded because its official release only provides a 1.5 kbps implementation at 16 kHz.

At 16 kHz and 1.1 kbps, the objective results in Table~\ref{16k} show that the lightweight VoCodec is highly competitive in LSD, STOI, and ViSQOL, approaching the heavyweight BigCodec despite much higher computational cost. This trend is consistent with the subjective MUSHRA results, where VoCodec ranks second, close to BigCodec and clearly ahead of StreamCodec.
Compared with StreamCodec, replacing the original RSVQ with the proposed voicing-driven quantizer and introducing a unidirectional LSTM yields clear quality gains without a significant increase in parameters, while preserving low latency, low complexity, and a lightweight design. The 48 kHz results in Table~\ref{48k} further confirm the effectiveness of VoCodec across sampling rates and datasets.

We further conducted ABX listening tests at 16 kHz. As shown in Fig.~\ref{ABX_1}, the subjective results are consistent with Table~\ref{16k}: at 1.1 kbps, VoCodec significantly outperforms most baselines ($p<0.05$) and remains comparable to BigCodec ($p>0.05$). This indicates that reducing the bitrate of unvoiced frames via voicing-driven quantization does not noticeably degrade perceptual quality.
We also compared VoCodec at 1.1 kbps with higher-bitrate baselines in order to evaluate the bitrate savings enabled by the proposed quantizer. Fig.~\ref{ABX_2} shows that VoCodec at 1.1 kbps achieves perceptual quality comparable to all compared codecs, including SQCodec, at 1.5 kbps ($p>0.05$), corresponding to a bitrate saving of 27\% (400 bps).
\vspace{-2mm}

\begin{table}[t]
\centering
    % \caption{Experimental results on decoded speech quality evaluations for StreamCodec, VoCodec and VoCodec-r at 1.1 kbps on the LibriTTS test set with a 16 kHz sampling rate, where $\text{LSD}_{v}$ and $\text{LSD}_{u}$ denotes the LSD of the voiced and unvoiced frames, respectively.}
    \caption{Experimental results on decoded speech quality evaluations for StreamCodec, VoCodec and VoCodec-r on LibriTTS (16 kHz) at 1.1 kbps.}
    \resizebox{=0.90\linewidth}{!}{
    \begin{tabular}{c | c c c c c}
        \hline
        
        \hline
& \multicolumn{5}{c}{Target Bitrate = 1.1 kbps}  \\
        \cline{2-6}
         & LSD $\downarrow$ & $\text{LSD}_{v}$ $\downarrow$ & $\text{LSD}_{u}$ $\downarrow$ & STOI $\uparrow$ & ViSQOL $\uparrow$\\
         \hline
         StreamCodec& 0.918 & 0.740  & \textbf{0.620} & 0.896 & 4.048 \\
         VoCodec & \textbf{0.896} & \textbf{0.700} & 0.645  & \textbf{0.916} &\textbf{4.115}\\
        VoCodec-r & 0.959 & 0.796 &   0.622  & 0.823 &   3.673  \\
         
        \hline
        
        \hline
    \end{tabular}}
\label{LSD}
\end{table}

\begin{figure}[t]
  \centering
  \includegraphics[width=0.90\linewidth]{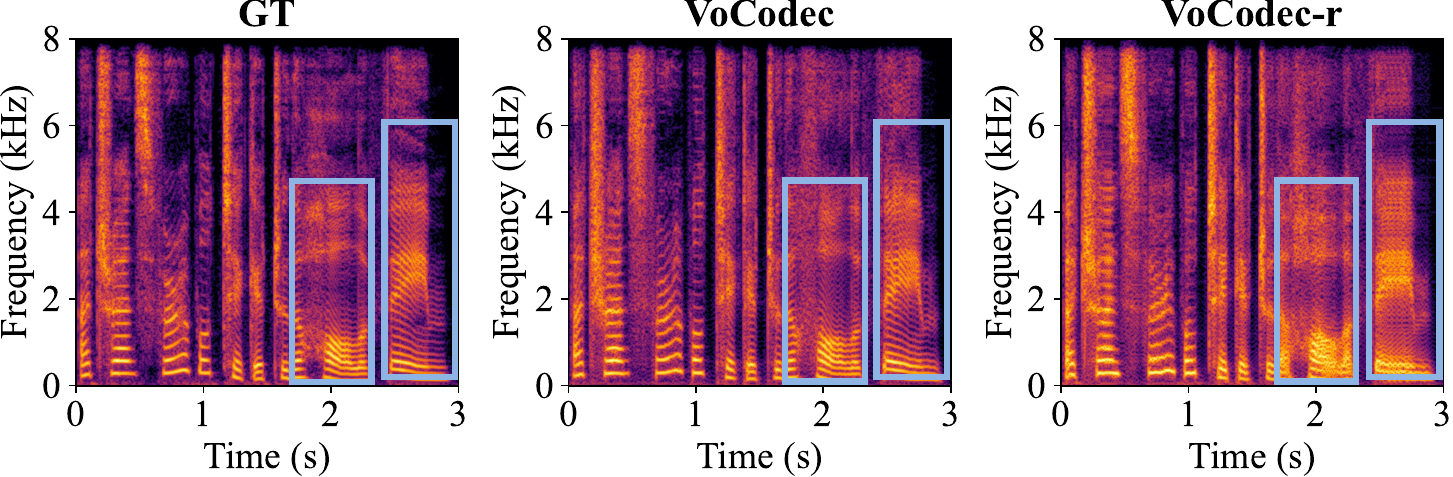}
  \caption{Comparison of speech spectrograms from the ground truth (GT) and the decoded speech of VoCodec and VoCodec-r.}
  \label{abla}
\end{figure}

\vspace{-1mm}
\subsection{Analysis and Discussion}
\vspace{-1mm}

In this section, we further conducted detailed analysis experiments for voiced and unvoiced frames respectively to verify the effectiveness of the speech coding approach based on voicing properties proposed in this paper. 
For convenience, the following experiments were conducted only at 1.1 kbps and a 16 kHz sampling rate. 

%\subsubsection{Experiments for Voiced and Unvoiced Sounds}

% \subsubsection{Separate Evaluation for Voiced and Unvoiced Frames}
\noindent\textbf{Separate Evaluation for Voiced and Unvoiced Frames.}
We measured LSD for voiced and unvoiced frames as $\text{LSD}_{v}$ and $\text{LSD}_{u}$ separately. 
Table \ref{LSD} shows that VoCodec, which prioritizes quantization of voiced frames, outperformed StreamCodec in voiced-frame LSD but performed worse on unvoiced frames.
However, subjective results in Fig. \ref{ABX_1} indicate that VoCodec sounded significantly better than StreamCodec. 
These findings suggest that the degraded objective performance on unvoiced frames has minimal impact on perceptual quality, consistent with the effectiveness of VoCodec’s voicing-driven quantization strategy.

% \subsubsection{Analysis of the Impact of Reversing Quantization Strategies Between Voiced and Unvoiced Frames}
\noindent\textbf{Analysis of the Impact of Reversing Quantization Strategies Between Voiced and Unvoiced Frames.}
% To further validate the effectiveness of the voicing-driven quantizer, we reversed the quantization strategies, i.e., using RSVQ for unvoiced frames and a single SQ for voiced frames (denoted as VoCodec-r). 
% By comparing VoCodec and VoCodec-r in Table \ref{LSD}, it can be observed that reversing the quantization strategies for voiced and unvoiced frames leads to a significant degradation in all metrics, including LSD, STOI, and ViSQOL. 
% Although applying a complex quantization strategy to unvoiced frames results in a noticeable improvement in their LSD, it does not contribute to speech intelligibility or overall perceptual quality. 
% Fig. \ref{abla} also presents the spectrograms of VoCodec and VoCodec-r. 
% As highlighted by the blue box in Fig. \ref{abla}, it is clearly observed that the voiced frame harmonics in the speech generated by VoCodec-r suffer from severe distortion. 
% These observations further demonstrate the necessity of employing a more complex quantization strategy for voiced frames than for unvoiced ones, thereby validating the effectiveness of the voicing-driven quantization approach introduced in VoCodec.
To further validate the effectiveness of the voicing-driven quantizer, we reversed the quantization strategies: using RSVQ for unvoiced frames and single SQ for voiced frames (denoted VoCodec-r).  
Table \ref{LSD} compares VoCodec and VoCodec-r, showing reversed quantization for voiced/unvoiced frames significantly degrades most metrics, including, LSD, $\text{LSD}_{v}$, STOI and ViSQOL. 
While complex quantization for unvoiced frames improves their LSD, i.e., $\text{LSD}_{u}$, it offers no gain in speech intelligibility or overall perceptual quality.  
Fig. \ref{abla} also presents their spectrograms, with the blue box highlighting severe distortion in the voiced-frame harmonics of VoCodec-r.
These results confirm the need for more complex quantization of voiced frames and the effectiveness of VoCodec’s voicing-driven approach.

\section{Conclusion}
\label{sec: Conclusion}

% In this paper, we present VoCodec, a streamable neural speech codec designed for low bitrate scenarios. 
% A novel voicing-driven quantization strategy is proposed as a key component of VoCodec. It allocates bitrates based on the voiced/unvoiced characteristics of speech, effectively reducing bitrate while maintaining high perceptual quality.
% Experimental results confirm that VoCodec achieves higher coding quality than advanced baseline neural speech codecs under lower bitrate conditions, while maintaining competitive computational and model complexity. This demonstrates its ability to save bitrate and improve compression efficiency.
% Extending VoCodec to the coding of other types of audio data will be the focus of our future work.

This paper presents VoCodec, a streamable neural speech codec designed for low-bitrate scenarios. 
Its key innovation is a voicing-driven quantization strategy that allocates bitrates based on speech's voiced/unvoiced characteristics, effectively reducing bitrate while preserving high perceptual quality. 
Experimental results show VoCodec outperforms advanced baseline neural speech codecs in coding quality at lower bitrates, with competitive computational and model complexity, validating its ability to save bitrate and improve compression efficiency. 
Extending VoCodec to the coding of other types of audio data will be the focus of our future work.

% This paper presents VoCodec, a streamable neural speech codec for low bitrates. 
% Its key innovation is a voicing-driven quantization to allocate bits across voiced/unvoiced segments, preserving perceptual quality at reduced rates. 
% Experiments show VoCodec outperforms advanced baseline neural speech codecs in coding quality at lower bitrates, with competitive computational and model complexity. 
% Extending VoCodec to the coding of other types of audio data will be the focus of our future work.

\section{Acknowledgments}
This work was supported by the National Natural Science Foundation of China under Grant No. 62301521.

\section{Generative AI Use Disclosure}
During the preparation of this manuscript, the authors used ChatGPT 5.2 to polish the language and improve the flow of the text. After using this tool, the authors reviewed and edited the content as needed and take full responsibility for the final version of the manuscript.

\bibliographystyle{IEEEtran}
\bibliography{mybib}

\end{document}